# Behavioural Change Support Intelligent Transportation Applications

Workshop at ITSC 2017 by Efthimios Bothos, Babis Magoutas, Brian Caulfield, Athena Tsirimpa, Maria Kamargianni, Panagiotis Georgakis and Gregoris Mentzas.

*Abstract*-This workshop invites researchers and practitioners to participate in exploring behavioral change support intelligent transportation applications. We welcome submissions that explore intelligent transportation systems (ITS), which interact with travelers in order to persuade them or nudge them towards sustainable transportation behaviors and decisions. Emerging opportunities including the use of data and information generated by ITS and users' mobile devices in order to render personalized, contextualized and timely transport behavioral change interventions are in our focus. We invite submissions and ideas from domains of ITS including, but not limited to, multi modal journey planners, advanced traveler information systems and in-vehicle systems. The expected outcome will be a deeper understanding on the challenges and future research directions with respect to behavioral change support through ITS.

## I. SCOPE AND GOALS

Behavioral change support (also known as persuasive technology) is commonly defined as technology designed to support positive behavioral change through persuasion and social influence, but not through coercion [1]. Related systems implement principles from social psychology to digital interfaces with the goal of motivating a user to achieve a target behavior or goal. Examples of such principles include incentives and rewards, social comparison, gamification and persuasive elements in computer interfaces.

The emergence of a multitude of ITS data and the pervasive use of technological systems in our everyday transportation (including smartphones, smart cards, in-vehicle smart systems) as well as related applications such as multimodal journey planners, community based traffic and navigation apps, is leading to a new generation of behavioral change support ITS technologies that can nudge travelers towards sustainable decisions through hyper personalized, context aware and timely interventions.

For example the analysis of ITS data capturing user actions can anticipate future behavior [2] and provide insights on behavioral patterns in novel contexts. In such settings traffic demand management approaches [3] can be utilized to dynamically manage, control and influence travel choices, traffic and facilities in an integrative framework. Such approaches are highly efficient as they enable transport agencies to leverage the existing investments so that a more efficient and effective system is achieved and the service life of existing capital investments is extended [3]. Moreover, traffic, mobile phone and social media data (e.g. from twitter) can reveal how humans move across cities, be used to assess the impact of online social networks (OSN) on individuals travel patterns [4], predict future mobility behavior (e.g. destination and time of day, mode, route) and provide indications of travel patterns, even predicting traffic incidents. Such information can be leveraged by behavioral change support ITS technologies in order to nudge and direct users to decisions that improves the overall performance and efficiency of the transportation system. For instance, indications of rising traffic can inform a personalized persuasive travel advisor app to properly nudge car users to shift their trip to a different time slot or opt for alternative transportation modes (e.g. public transportation). The app can present the savings users will be able to reach in terms of time and $CO_2$ emissions if they shift their trip to a different time slot when traffic will be lower, if such a decision suites their needs, or by selecting public transportation to reach their destination. Other prominent applications of persuasive ITS concern in-vehicle systems which aim to increase sustainable behavior through e.g. good driving habits for fuel and energy efficiency [5].

Behavioral change support and persuasive design strategies take various forms [6]. They can be strong and provide information that shows the extent to which a user's behavior is or is not sustainable, or they can be passive and present more general information on impact effects (e.g. consumption) contextualized within the topic of sustainable transportation. Other strategies can actively support decisions and guide users to follow particular behavior patterns e.g. [7], [8]. Moreover, approaches can be implicit and un-consciously persuade the users through e.g. ambient technologies [9].

Recent results on behavioral change support and persuasive technologies show that personalization, context awareness and proactivity can greatly affect sustainable interventions [10]. This workshop intends to explore the opportunities provided by personal big data [11] and big data generated by ITS and user mobile devices in order to render the behavioral change support interventions personalized, contextualized and timely.

### A. Workshop's Goal

The goal of this workshop is to discuss opportunities and challenges for behavioral change support ITS technologies from various perspectives, including considerations of adaptive and personalized user interfaces and interaction techniques, contextualized feedback approaches, choice architecture, decision support approaches and proactive interventions that maximize the impact of behavioral change support interventions. Further, we plan to discuss our insights in the light of future transportation solutions (e.g. multimodality, sharing systems, electric vehicles, on-demand services etc.). The overall objective is to identify and summarize recent challenges in the design and development of behavioral change support ITS technologies. The following list provides an overview of topics relevant to the workshop (although non-exhaustive).

- Applying behavioral change in transportation: methodological aspects
- Persuasive technologies and choice architecture for transport applications
- Gamification approaches and serious games, incentives for behavioral change and persuasion
- Context-aware ITS for traveler decisions
- Behavioral and personality analysis for persuasion in ITS
- Timing of traveler persuasive interventions
- Emerging possibilities for ITS persuasion through new technologies
- In-vehicle behavioral change support ITS applications
- Transport Safety and ITS behavioral change applications
- Behavioral data mining through ITS (including smart card and floating car data) for persuasive interventions
- Behavioral change interventions for sustainable transport
- Evaluation of behavioral change support ITS applications
- Ethical considerations in transport persuasion
- Social interactions and behavioral change in transportation
- Social mining for supporting ITS applications

*B. Workshop's Format*

We intend to host a half-day workshop comprising of a keynote speech, short participant talks and focused discussion groups. The workshop will start with a general introduction followed by a brief keynote motivating the topic at hand. We will then proceed with paper presentations: participants will have 15 minutes to present their work, followed by 5 minutes discussion and questions. A discussion session and group work will follow to identify challenges and opportunities for behavioral change support ITS technologies. We will conclude the workshop by summarizing the key insights and plan follow-up activities. All papers and presentations as well as a summary of the workshop and the key findings will be published on the workshop website, whereas our goal is to put in place the first steps for a journal special issue, (e.g. at the International Journal of Intelligent Transportation Systems Research).

The audience for this workshop comprises of the multidisciplinary cross section of transportation, HCI and persuasive technology, including researchers and practitioners who are active in areas like behavioral change support systems, context-aware ITS, travel behavioral analysis, personalization and recommendation systems, sustainable HCI. We expect a good mix of industrial and academic participants, which will lead to lively and insightful discussions.

*C. Outcomes and Future Directions*

The contributions will provide a deeper understanding on the challenges and future research directions with respect to behavioral change support and persuasive ITS. Further, the discussions at the workshop will offer an opportunity to understand new possibilities for behavioral change support and persuasive ITS technology as it is currently constituted and to list potential limits. We expect that this list will contain a brief summary of the state of the art as well as research directions/ contributions and collaborations necessary to enable the creation of a next generation of behavioral change support and persuasive ITS technologies.

*D. Organizers*

**Dr. Efthimios Bothos** is a senior researcher at the Institute of Information and Computer and Communication Systems of the National and Technical University of Athens (NTUA). His current research interests include Persuasive technologies for transport applications, Information Aggregation Systems, Recommendation and Personalization Systems. Email: mpthim@mail.ntua.gr.

**Dr. Babis Magoutas** is a post-doctoral researcher at the School of Electrical and Computer Engineering of the National Technical University of Athens (NTUA) His current research interests include persuasive technologies, recommender systesm, information personalization, information systems evaluation and situation awareness. Email: elbabmag@mail.ntua.gr.

**Dr. Brian Caulfield** is an associate Professor of the Department of Civil, Structural and Environmental Engineering at the Trinity College Dublin. Dr Caulfield has embarked on research programs addressing global issues such as the environmental impacts of transport and methods to reduce the carbon impacts of transport. Dr Caulfield has published over 100 papers in these areas in high impact international journals and international conferences, Email: Brian.Caulfield@tcd.ie.

**Dr. Athena Tsirimpa** is a senior researcher at the Transportation and Decision Making Laboratory of the University of the Aegean. Her current research interests include travel behavioral analysis, behavioral change, emotions, personalization and traveler information systems. Email: atsirimpa@aegean.gr.

**Dr. Maria Kamargianni** is a lecturer at the University College London (UCL). Her areas of research include travel behavior modeling, transportation systems analysis, mobility as a service (MaaS), new mobility services and business models, active transportation (walking and cycling) modelling, social networking and traveling, demand analysis, market research and econometrics. She has worked on various transportation projects in Europe and worldwide and is involved in campaigns for active transport, green transportation and road safety. Email: m.kamargianni@ucl.ac.uk.

**Dr. Panagiotis Georgakis** is a reader in transportation systems at the Sustainable Transport Research Group of the University of Wolverhampton. His areas of interest are ITS system design and development, in-vehicle networking and integration, AI techniques for urban planning, multi-criteria evaluation for logistical operations and others. Email: p.georgakis@wlv.ac.uk

**Dr. Gregoris Mentzas** is Professor of Management Information Systems at the School of Electrical and Computer Engineering of the National Technical University of Athens (NTUA) and Director of the Information Management Unit (IMU), a multidisciplinary research unit at the University. His area of expertise is information technology management and his research concerns knowledge management, semantic web and e-service technologies in e-government and e-business settings. Email: gmentzas@mail.ntua.gr.

ACKNOWLEDGMENT

The workshop is partly supported by the EU project OPTIMUM (H2020 grant agreement no. 636160-2).